\documentclass{ws-p9-75x6-50}
\input psfig.tex
\begin{document}

\title{State-or-the-art Accretion and Wind Solutions Around Black Holes}

\author{Sandip K. Chakrabarti}

\address{S.N. Bose National Centre for Basic Sciences, JD-Block, Salt Lake, Calcutta
700098\\E-mail: chakraba@boson.bose.res.in}

\maketitle

\abstracts{We report all possible basic solutions of an advective disk model and use them
to explain major observations from black hole candidates.}

\noindent Proceedings of 9th Marcel grossman Meeting in Rome (Ed. R. Ruffini)

Advective flow solutions are around over ten years now\cite{chak90}, but because of complexity
they are difficult to use by observers. These solutions show complex behaviours, both spatially and temporally.
On the spatial front, it predicts that (a) Matter must be supersonic and sub-Keplerian just outside the
horizon. (b) Matter takes very little time for the infall and therefore
angular momentum must be roughly conserved. Hence the centrifugal barrier supported 
boundary layer (CENBOL) would be produced. (c) Steady CENBOL which is a few tens of Schwarzschild radii
big, decides the nature of the emitted spectrum and the outflow rate in the jets.
(d) Flows with a viscosity higher than a critical value do not have CENBOL. They start from
Keplerian disks and enter into black holes through the inner sonic point.
On the temporal front, it predicts that (a) In the absence of a steady solution,
the CENBOL oscillates periodically when viscosity is small enough. (b) Quasi-Periodic Oscillation (QPO) of 
the X-rays should be produced because of this oscillation of the CENBOL. Complete classification of these
solutions have been discussed in various contexts in recent years\cite{skc98,ijprev,hehl,pes1,samar,granada1}.

Figure 1 gives the `alphabets' or the building blocks of the advective disk 
solution\cite{granada1}. Typical solutions
(Mach number along Y-axis and logarithmic radial distance along X-axis) are shown in Figs. (A-J) and schematic
flow behaviours are shown in Figs. (a-j) respectively. In A (low energy)
and F (high energy) solutions have only saddle type sonic points. The outflow and wind solutions
are quasi-conical (a,f) since angular momentum is low. In C (shocks in accretion) and D (shocks
in winds) there are three sonic points and Rankine-Hugoniot conditions  are satisfied to form 
steady shocks. A large region of the parameter space produces these flows (c,d). In B (oscillating shocks in
accretion) and E (oscillating shocks in jets) Rankine-Hugoniot conditions are not satisfied, but there are
three sonic points. These solutions are time dependent and therefore important for 
observational point of view (b,e). In G and H, inviscid solutions are incomplete  and 
flow would be noisy (g, h). With viscosity, solutions in C become those of I and J
depending on whether viscosity is below critical (I,i) or above critical (J, j). A realistic flow
around a black hole must be combinations of these ingredients. For instance, a flows with high viscosity
may be (I,i) type in the equatorial plane and those with low viscosity may be (J,j) type away from the
plane, thus creating a two-component advective flow\cite{ct95}. Similarly, oscillating shocks
satisfactory explains QPOs. CENBOL free solutions are known to produce lesser outflows as in 
soft states. Thus advective solutions, though compromise with simplicity, create a unique
paradigm in which one could work without changing models from one observation to another.

\begin{figure}
\vbox{
\vskip -1.0cm
\hskip 0.0cm
\centerline{
\psfig{figure=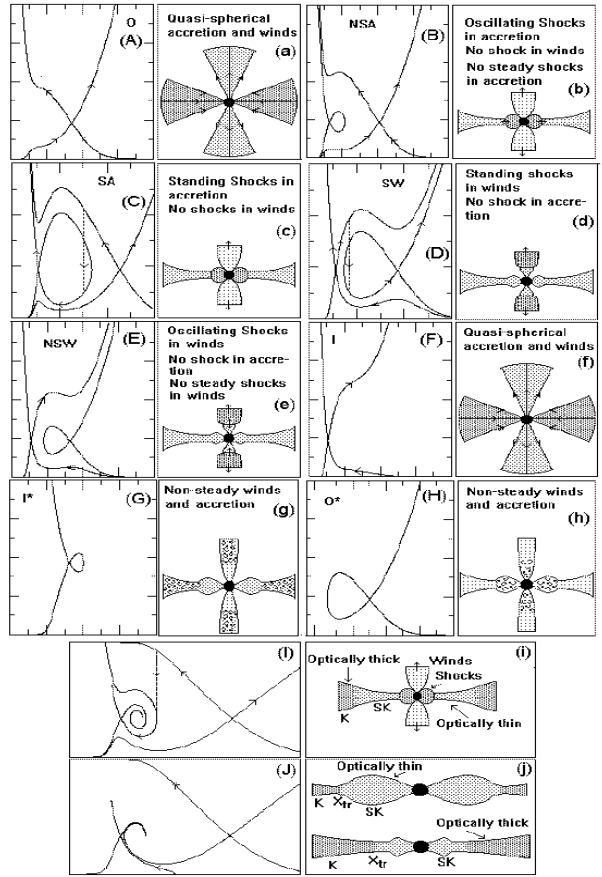,height=12truecm,width=8truecm}}}
\vspace{0.0cm}
\caption[] {Classification of all types of advective flows in inviscid (A-H) flows
and some of the viscous flows (I-J). Schematic nature of the flows
are shown in a-j. See text for details.}
\end{figure}

\end{document}